\documentclass{article}
\usepackage{spconf,amsmath,graphicx}

\usepackage{cite}
\usepackage{amsmath,amssymb,amsfonts}
\usepackage{graphicx}
\usepackage{textcomp}
\usepackage{xcolor}
\usepackage[acronym,shortcuts]{glossaries}
\usepackage{bm}
\usepackage{caption}
\usepackage{subcaption}
\usepackage{relsize}
\usepackage{soul}
\usepackage{algorithm, algpseudocode}
\usepackage{dblfloatfix}

\def\BibTeX{{\rm B\kern-.05em{\sc i\kern-.025em b}\kern-.08em
T\kern-.1667em\lower.7ex\hbox{E}\kern-.125emX}}

\newcommand{\trans}[0]{^{\mathsf{T}}}

\newcommand{\herm}[0]{^{\mathsf{H}}}

\newacronym{OTFS}{OTFS}{Orthogonal Time Frequency Space}
\newacronym{CRLB}{CRLB}{Cram{\`e}r-Rao lower bound}
\newacronym{AoA}{AoA}{angle-of-arrival}
\newacronym{SNR}{SNR}{signal-to-noise ratio}
\newacronym{ML}{ML}{maximum likelihood}
\newacronym{MIMO}{MIMO}{multiple-input multiple-output}
\newacronym{SISO}{SISO}{single-input single-output}
\newacronym{MUSIC}{MUSIC}{multiple signal classification}
\newacronym{ROOT-MUSIC}{ROOT-MUSIC}{ROOT multiple signal classification}
\newacronym{JCAS}{JCAS}{joint communications and sensing}
\newacronym{JCR}{JCR}{joint communications and radar}
\newacronym{ISAC}{ISAC}{integrated sensing and communications}
\newacronym{3D}{3D}{three-dimensional}
\newacronym{2D}{2D}{two-dimensional}
\newacronym{ROI}{ROI}{region of interest}
\newacronym{BF}{BF}{beamforming}
\newacronym{mmWave}{mmWave}{milimeter-wave}
\newacronym{MF}{MF}{matched-filter}
\newacronym{DD}{DD}{delay-Doppler}
\newacronym{SotA}{SotA}{State-of-the-Art}
\newacronym{ULA}{ULA}{uniform linear array}
\newacronym{QAM}{QAM}{quadrature amplitude modulation}
\newacronym{ISFFT}{ISFFT}{inverse symplectic finite Fourier transform}
\newacronym{SFFT}{SFFT}{symplectic finite Fourier transform}
\newacronym{ISI}{ISI}{inter-symbol interference}
\newacronym{AWGN}{AWGN}{additive white Gaussian noise}
\newacronym{MSE}{MSE}{mean-squared-error}
\newacronym{LMMSE}{LMMSE}{linear minimum mean square error}
\newacronym{RMSE}{RMSE}{root mean square error}
\newacronym{ESPRIT}{ESPRIT}{estimation of signal parameters via rotational invariant techniques}
\newacronym{OFDM}{OFDM}{Orthogonal Frequency Division Multiplexing}

\title{Fast and Efficient Sequential Radar Parameter\\ Estimation in MIMO-OTFS Systems\\[-2ex]}
\name{Kuranage Roche Rayan Ranasinghe$^\dag$, Hyeon Seok~Rou$^\dag$, Giuseppe~Thadeu~Freitas~de~Abreu$^\dag$\thanks{K.~R.~R.~N.~Ranasinghe, H.~S.~Rou and G.~T.~F.~Abreu are with the School of Computer Science and  Engineering, Constructor University (\textit{previously Jacobs University Bremen}), Campus Ring 1, 28759, Bremen, Germany (emails: [kranasinghe, hrou, gabreu]@constructor.university).}}
\address{\vspace{-3ex}$^\dag$School of Computer Science and Engineering, Constructor University, Germany}
\begin{document}

%
\maketitle
\begin{abstract}
We consider the estimation of \ac{3D} radar parameters, namely, bearing or \ac{AoA}, delay or range, and Doppler shift velocity, under a mono-static \ac{MIMO} \ac{JCR} system based on \ac{OTFS} signals.
In particular, we propose a novel two-step algorithm to estimate the three radar parameters sequentially, where the \ac{AoA} is obtained first, followed by the estimation of range and velocity via a reduced \ac{2D} grid \ac{ML} search in the \ac{DD} domain. 
Besides the resulting lower complexity, the decoupling of \ac{AoA} and \ac{DD} estimation enables the incorporation of an \ac{LMMSE} procedure in the \ac{ML} estimation of range and velocity, which are found to significantly outperform \ac{SotA} alternatives and approach the fundamental limits of the \ac{CRLB} and search grid resolution.
\end{abstract}
\begin{keywords}
\Ac{JCR}, \ac{OTFS}, \ac{MIMO}, \ac{AoA}, \ac{DD}, \ac{LMMSE}, Root \ac{MUSIC}, low-complexity.
\end{keywords}

\glsresetall

\vspace{-2ex}
\section{Introduction}
\label{sec:intro}
\vspace{-1ex}

Besides its inherent robustness to mobility, \ac{OTFS} \cite{Hadani_WCNC_2017} has received great attention since its inception due to its ability to enable both communication and radar sensing applications \cite{Raviteja_RadarConf_2019,Gaudio_TWC_2020,Gaudio_RadarConf20_2020, Bondre_RadarConf22_2022,Muppaneni_CommLetters_2023}, via the exploitation of its \ac{DD} channel  structure.

Various radar techniques have been proposed to extract target information (delay and Doppler parameters) under \ac{SISO} set-ups.
One of the first attempts in \cite{Raviteja_RadarConf_2019}, utilizes a simple \ac{MF}-based approach to detect a single target. 
Later, other methods considering more realistic scenarios were proposed \cite{Gaudio_TWC_2020}, where a \ac{ML}-based method was used to iteratively estimate both delay and Doppler parameters in a multi-target situation. 

It was identified, however, that the complexity of the \ac{ML}-based methods is prohibitive, especially with high precision \cite{Muppaneni_CommLetters_2023}, since a full grid-search is required across the vicinity of the \ac{DD} plane.
This is further exacerbated when evolving from \ac{SISO} to \ac{MIMO} radar systems, which is considered a key technology for \ac{JCAS}, due to its ability to distinguish multiple targets via the additional spatial dimension \cite{Li-J_Book_2009}, whose diversity is exploited in communication systems to improve the performance and robustness of modulation schemes. 

Coupled with \ac{BF} design and the calibration of co-located transmit and receive antenna arrays, \ac{MIMO} radar is an extremely effective way to cover large angular sectors for target detection and radar parameter estimation.
The performance gains of \ac{MIMO}-\ac{OTFS} systems come, however, at the price of a third signal domain -- namely, the \ac{AoA} domain -- such that the radar parameter becomes a \ac{3D} search problem
of overwhelming complexity, as discussed in \cite{Gaudio_RadarConf20_2020}.

In light of the above, we contribute to improving the performance of \ac{3D} radar parameter estimation using \ac{MIMO}-\ac{OTFS} signals, with a reduction of complexity as a bonus. 
To that end, a novel two-step algorithm is proposed whereby \ac{AoA} and \ac{DD} estimation are decoupled, such that in addition to essentially converting the original \ac{3D} \ac{MIMO}-\ac{OTFS} problem to \ac{2D} search of complexity comparable to that of the earlier \ac{SISO} case \cite{Muppaneni_CommLetters_2023}, the incorporation of a \ac{LMMSE} procedure into the formulation of the \ac{DD} search is enabled, resulting in substantial gains in performance.

\begin{figure*}[b!]
\hrulefill
\setcounter{equation}{7}
\normalsize
\begin{equation}
\setlength{\abovedisplayskip}{1pt}
\setlength{\belowdisplayskip}{0.1pt}
\label{eq:otfs_channel_tf}
\mathbf{h}_{n,m}[n',m'] = \sum_{p=1}^{P} h'_p \cdot \mathbf{b}(\phi_p) \mathbf{a}\herm(\phi_p) \mathbf{f}_{\mathrm{BF}} \cdot C_{g_{\text{tx}},g_{\text{rx}}} \big((n-n')T - \tau_p , (m-m') \Delta f - \nu_p \big) \cdot e^{j2\pi n' T\nu_p} \cdot e^{j2\pi m\Delta f\tau_p} \in \mathbb{C}
\vspace{-1ex}
\end{equation}
\setcounter{equation}{10}
\begin{equation}
\label{eq:otfs_psi}
{\varPsi}^p_{k,k'}[\ell,\ell'] \approx  \frac{1}{NM} \cdot \frac{1 - e^{j2\pi (k' - k + \nu_p NT)}}{1 - e^{j2\pi \frac{(k' - k + \nu_p NT)}{N}}} \cdot \frac{1 - e^{j2\pi (\ell' - \ell + \tau_p M\Delta f)}}{1 - e^{j2\pi \frac{(\ell' - \ell + \tau_p M\Delta f)}{M}}} e^{-j2\pi \nu_p \frac{\ell'}{M\Delta f}} 
\begin{cases}
1,& \text{if } \ell' \in \mathcal{L}_{ICI}\\
e^{-j2\pi (\frac{k'}{N} + \nu_p T)},& \text{if } \ell' \in \mathcal{L}_{ISI}
\end{cases}
\end{equation}
\setcounter{equation}{0}
\end{figure*}

\vspace{-3ex}
\section{System Model}
\label{sec:system_model}
\vspace{-2ex}
\subsection{MIMO-OTFS Channel Model}
\label{subsec:otfs_channel_model}
\vspace{-1ex}

Consider a \ac{ULA} \ac{MIMO} radar  equipped with $N_\mathrm{a}$ antennas, operating in full-duplex mode as in \cite{Gaudio_RadarConf20_2020}, at a carrier frequency of $f_\mathrm{c} \;\mathrm{Hz}$ with channel bandwidth $B \; \mathrm{Hz}$. 

It is assumed that a point target model is adopted for the system, such that each target can be modeled through its line-of-sight (LoS) path only \cite{Nguyen_ICASSP_2017} and thus represented by a unique single tap in the \ac{DD} channel corresponding to the round-trip of the signal.
Then, the corresponding time-frequency selective \ac{OTFS} channel with $P$ taps \cite{Gaudio_RadarConf20_2020,Vitetta_Book_2013} can be described as 
\begin{equation}
\setlength{\abovedisplayskip}{1pt}
\setlength{\belowdisplayskip}{0.2pt}
\!\mathbf{H}(t,\tau)\!=\!\!\sum_{p=1}^P h_p \! \cdot\!   \delta(\tau \! - \! \tau_p)  \!\cdot\!  e^{j2\pi \nu_p t} \! \cdot \! \mathbf{b}(\!\phi_p \!) \mathbf{a}\herm(\!\phi_p \!) \in \mathbb{C}^{N_a \times N_a}\!,\!\!\!\!\!
\label{eq:otfs_channel_eq}
\end{equation}
where $h_p \in \mathbb{C}$ with $p \in \{1,\cdots,P\}$ is the complex channel gain including the pathloss component;
$\tau_p$ and $\nu_p$ are the round-trip delay and Doppler-shift of the $p$-th target given by
\begin{equation}
\setlength{\abovedisplayskip}{2.5pt}
\setlength{\belowdisplayskip}{2.5pt}
\tau_p \triangleq \frac{2r_p}{c} \in \mathbb{R}, ~~\text{and}~~
\nu_p \triangleq \frac{2v_p f_\mathrm{c}}{c} \in \mathbb{R},
\label{eq:roundtrip_parameters}
\end{equation} 
with $r_p \in \mathbb{R}$, $v_p \in \mathbb{R}$ denoting the range and velocities of the $p$-th target, respectively, and $c$ denoting the velocity of light at $299,792,458 \; \mathrm{m/s}$;
$\phi_p \in [-\frac{\pi}{2},\frac{\pi}{2}]$ is the \ac{ULA} steering angle of the $p$-th target;
and $\mathbf{a}(\phi) = [a_1(\phi), \cdots, a_{N_\mathrm{a}}\!(\phi)]\trans$ $\in$ $\mathbb{C}^{N_\mathrm{a} \times 1}$ and $\mathbf{b}(\phi) = [b_1(\phi), \cdots, b_{N_\mathrm{a}}\!(\phi)]\trans$ $\in$ $\mathbb{C}^{N_\mathrm{a} \times 1}$ are respectively the transmit and receive steering vectors with elements
\begin{equation}
\setlength{\abovedisplayskip}{2pt}
\setlength{\belowdisplayskip}{0.2pt}
a_n(\phi) = b_n(\phi) = e^{j(n-1)\pi \sin(\phi)} \in \mathbb{C}.
\label{eq:ULA_resp_vect}   
\end{equation}

\vspace{-3ex}
\subsection{OTFS Transmit Signal Model}
\label{subsec:otfs_transmit_signal_model}
\vspace{-1ex}

The \ac{OTFS} signal structure considered is classic \cite{Gaudio_RadarConf20_2020, Gaudio_TWC_2020}, with $M$ subcarriers of bandwidth $\Delta f \triangleq B/M$, each conveying $N$ modulated symbols taken from an arbitrary  \ac{QAM} constellation $\mathcal{X}$ with symbol duration $T \triangleq 1/{\Delta f}$, such that the total \ac{OTFS} frame duration is $NT$.

In other words, the modulated symbols $x_{k,\ell} \in \mathbb{C}$, represented by the \ac{DD} domain symbol matrix $\mathbf{X} \in \mathbb{C}^{N \times M}$, with $k \in \{0,\cdots,N-1\}$ and $\ell \in \{0,\cdots,M-1\}$, are arranged in the \ac{DD} domain grid $\Gamma$, with a corresponding delay of $k/NT$ seconds and a Doppler of $\ell/M\Delta f$ $\mathrm{Hz}$.
The transmitter performs the \ac{ISFFT} on the \ac{DD} grid to convert the data into the dual (time-frequency) domain, whose samples are given by
\vspace{-1.5ex}
\begin{equation}
X[n,m] = \sum_{k=0}^{N-1} \sum_{l=0}^{M-1} x_{k,\ell} 
\cdot e^{j2\pi (\frac{nk}{N} - \frac{m\ell}{M})} \in \mathbb{C},
\label{eq:otfs_isfft}  
\vspace{-1ex}
\end{equation}
with $n = \{0,\dots,N-1\}$ and $m = \{0,\dots,M-1\}$, obeying the average power constraint $\mathbb{E}[|X[n,m]|^2] \leq P_{\text{avg}}/N_a$.

Utilizing the time-frequency data samples $X[n,m]$ from eq. \eqref{eq:otfs_isfft}, the actual transmitted continuous time domain signal is obtained by the Heisenberg transform,
\begin{equation}
\setlength{\abovedisplayskip}{2pt}
\setlength{\belowdisplayskip}{2pt}
s(t) = \sum_{n=0}^{N-1} \sum_{m=0}^{M-1} X[n,m] \cdot g_{\mathrm{tx}}(t - nT) \cdot e^{j2\pi m\Delta f(t-nT)},
\label{eq:otfs_s_t}    
\end{equation}
where $g_{\mathrm{tx}}(t)$ is a specific pulse-shaping function.

For the sake of simplicity, we here follow \cite{Gaudio_RadarConf20_2020} and consider that the \ac{MIMO} setup is employed for the radar application only\footnote{The extension to the case when $s(t)$ is generalized to a vector carrying multiple orthogonal data streams is also possible and was considered in \cite{Gaudio_PhD_2022}. We shall address such a case also in a journal version of this article.} \cite{Fortunati_TSP_2020}, such that the same symbol $s(t)$ is transmitted from all $N_a$ antennas simultaneously, subjected to the beamforming vector $\mathbf{f}_{\mathrm{BF}} \in \mathbb{C}^{N_\mathrm{a} \times 1}$ \cite{Friedlander_TAES_2012}.

\vspace{-3ex}
\subsection{OTFS Received Signal Model}
\label{subsec:otfs_received_signal_model}
\vspace{-1ex}

After transmission of the \ac{MIMO}-\ac{OTFS} signal over the time-delay channel specified in equation \eqref{eq:otfs_channel_eq}, the continuous time-domain received signal at the $N_a$ antennas is given by
\begin{equation}
\setlength{\abovedisplayskip}{2pt}
\setlength{\belowdisplayskip}{2pt}
\mathbf{r}(t) = \sum_{p=1}^{P} h_p \cdot e^{j2\pi \nu_p t} \cdot s(t - \tau_p)  \cdot \mathbf{b}(\phi_p)\mathbf{a}\herm(\phi_p) \mathbf{f}_{\text{BF}},
\label{eq:otfs_r_t}
\end{equation}
where the received noise is neglected for the time being.

By applying a receive pulse-shaping function $g_{\mathrm{rx}}(t)$ and sampling the signal in time-frequency with rates $t = nT$ and $f = m\Delta f$ in accordance to typical \ac{OTFS} signal processing methods, the received data samples are obtained as
\begin{equation}
\setlength{\abovedisplayskip}{2pt}
\setlength{\belowdisplayskip}{2pt}
Y[n,m] = \sum_{n'=0}^{N-1} \sum_{m'=0}^{M-1} X[n',m'] \cdot \mathbf{h}_{n,m}[n',m'],
\label{eq:otfs_y_tf}
\end{equation}
where the time-frequency domain channel vector ${\mathbf{h}_{n,m}[n',m']}\\ \in \mathbb{C}$ is given in eq. \eqref{eq:otfs_channel_tf}, with $C_{u,v}(\tau,\nu) \triangleq \int_{\infty}^\infty u(s) v^* (s - \tau) e^{-j2\pi \nu s} \; ds$ denoting the cross-ambiguity function between arbitrary $u(\cdot)$ and $v(\cdot)$ \cite{Gaudio_RadarConf20_2020}, and $h'_p \triangleq h_p \cdot e^{j2\pi \nu_p \tau_p} \in \mathbb{C}$ is the phase-rotated channel coefficient on $\nu_p$ and $\tau_p$.

Consequently, the received signal samples in time-freque-\\ncy domain are then converted back to the corresponding \ac{DD} domain via the \ac{SFFT} to yield the \ac{DD} samples $\mathbf{Y}[k,\ell] \in \mathbb{C}, \;\forall k, \ell$, as

\quad\vspace{-2ex}
\setcounter{equation}{8}
\setlength{\abovedisplayskip}{2pt}
\setlength{\belowdisplayskip}{2pt}
\begin{align}
\mathbf{Y}[k,\ell] &= \frac{1}{NM}\sum_{n = 0}^{N - 1} \sum_{m = 0}^{M - 1} {Y[n,m]} e^{j2\pi (\frac{m\ell}{M} - \frac{nk}{N})} \nonumber \\
& = \sum_{k'=0}^{N-1} \sum_{\ell'=0}^{M-1} x_{k',\ell'} \mathbf{g}_{k,k'}[\ell,\ell'] \in \mathbb{C},
\label{eq:otfs_y_k_l} 
\end{align}

\noindent where $\mathbf{g}_{k,k'}[\ell,\ell'] \in \mathbb{C}$ is the \ac{ISI} coefficient of the $[k',\ell']$-th \ac{DD} symbol as seen by the $[k,\ell]$-th sample, which is given by
\begin{equation}
\setlength{\abovedisplayskip}{2pt}
\setlength{\belowdisplayskip}{2pt}
\mathbf{g}_{k,k'}[\ell,\ell']\! = \!\!\sum_{p=1}^{P} h'_p \!\cdot\! \mathbf{b}(\phi_p)\mathbf{a}\herm (\phi_p) \mathbf{f}_{\mathrm{BF}} \!\cdot\! {\varPsi}_{k,k'}^p [\ell,\ell'] \in \mathbb{C},\!\!\!
\label{eq:otfs_isi_coeff}    
\end{equation}
where by approximating the cross-ambiguity function\footnotemark \,
as addressed in \cite{Gaudio_TWC_2020}, ${\varPsi}_{k,k'}^p [\ell,\ell']$ can be approximated as eq. \eqref{eq:otfs_psi}. 

From the above, it follows that the channel for the $p$-th target can be defined as
\begin{equation}
\setcounter{equation}{12}
\setlength{\abovedisplayskip}{2pt}
\setlength{\belowdisplayskip}{2pt}
\mathbf{G}_p(\tau_p , \nu_p , \phi_p) \triangleq \big(\mathbf{b}(\phi_p)\mathbf{a}\herm (\phi_p) \mathbf{f}_{\mathrm{BF}}\big) \otimes \mathbf{\Psi}^p \in \mathbb{C}^{N_\mathrm{a} NM\times NM}\!,
\label{eq:otfs_G}    
\end{equation}
where $\otimes$ denotes the Kronecker product, and $\mathbf{\Psi}^p \in \mathbb{C}^{NM \times NM}$ is a block matrix defined as
\begin{equation}
\boldsymbol{\Psi}^p \triangleq
\begin{bmatrix}
\boldsymbol{\Psi}^p_{1,1} & \!\!\!\!\!\cdots\!\!\!\!\! & \boldsymbol{\Psi}^p_{1,k'} & \!\!\!\!\!\cdots\!\!\!\!\! & \boldsymbol{\Psi}^p_{1,N} \\[-0.75ex]
\vdots & \!\!\!\!\!\ddots\!\!\!\!\! & \vdots & \!\!\!\!\!\ddots\!\!\!\!\! & \vdots  \\
\boldsymbol{\Psi}^p_{k,1} & \!\!\!\!\!\cdots\!\!\!\!\! & \boldsymbol{\Psi}^p_{k,k'} & \!\!\!\!\!\cdots\!\!\!\!\! & \boldsymbol{\Psi}^p_{k,N}  \\[-0.75ex]
\vdots & \!\!\!\!\!\ddots\!\!\!\!\! & \vdots & \!\!\!\!\!\ddots\!\!\!\!\! & \vdots  \\
\boldsymbol{\Psi}^p_{N,1} & \!\!\!\!\!\cdots\!\!\!\!\! & \boldsymbol{\Psi}^p_{N,k'} & \!\!\!\!\!\cdots\!\!\!\!\! & \boldsymbol{\Psi}^p_{N,N} \\
\end{bmatrix} \in \mathbb{C}^{NM \times NM}\!,\!\!\!\!\!\!
\end{equation}
where each block $\boldsymbol{\Psi}^p_{k,k'} \in \mathbb{C}^{M \times M}\!, \;\! \forall k,k'$ is a matrix with element at the $\ell$-th row and $\ell'$-th column is given by $\varPsi_{k,k'}[\ell,\ell']$ as described in equation \eqref{eq:otfs_psi}.

\footnotetext{Note that by assuming the pulse shaping functions $g_{\text{tx}}(t)$ and $g_{\text{rx}}(t)$ to be practical rectangular pulses of duration $T$ and amplitude $1/\sqrt{T}$, the formulation reduces to an equivalent Zak Transform \cite{Hong_Book_2022}.}  

Finally, by vectorizing the \ac{DD} domain symbol matrix $\mathbf{X} \in \mathbb{C}^{N \times M}$ of equation \eqref{eq:otfs_isfft} into $\mathbf{x} \in \mathbb{C}^{NM \times 1}$, the received signal matrix $\mathbf{Y} \in \mathbb{C}^{N \times M}$ of equation \eqref{eq:otfs_y_k_l} can also be expressed in the vectorized form
\vspace{-0.25ex}
\begin{equation}
\mathbf{y} = \sum_{p=1}^{P} \big(h'_p \!\cdot \mathbf{G}_p(\tau_p , \nu_p , \phi_p)\big) \mathbf{x} + \mathbf{w} \in \mathbb{C}^{N_\mathrm{a}NM \times 1}, 
\label{eq:otfs_y_DD}    
\vspace{-0.5ex}
\end{equation}
where $\mathbf{w} \sim \mathcal{CN}(\mathbf{0}, \sigma^2_w \mathbf{I}) \in \mathbb{C}^{N_\mathrm{a}NM \times 1}$ denotes the received \ac{AWGN} vector.

In light of the above, the received signal model sampled in \ac{DD} domain is concisely described in terms of four parameters per $p$-th target, \textit{i.e.,} complex channel coefficient $h_p$, round-trip delay $\tau_p$, Doppler-shift $\nu_p$, and \ac{AoA} $\phi_p$.

\vspace{-3ex}
\section{Radar Parameters Estimation Methods}
\label{sec:radar_est}
\vspace{-2ex}

Let $\boldsymbol{\theta}_p \triangleq \{h'_p, \tau_p, \nu_p, \phi_p \} \in \mathcal{T}$ be the parameters corresponding to a given $p$-th target, with $\mathcal{T}\! =\! \mathbb{C}\! \times\! \mathbb{R} \!\times\! \mathbb{R}\! \times\! [-\frac{\pi}{2},\frac{\pi}{2}]$. 
Since $\mathbf{x}$ is known at the mono-static \ac{MIMO}-\ac{OTFS} radar transmitter, the  parameter estimation problem turns to a joint search of  $4P$ parameters $\boldsymbol{\theta} \triangleq \{h'_1,\! \cdots\!, h'_P,  \tau_1,\! \cdots\!, \tau_P, \nu_1,\!\cdots\!,\nu_P, \phi_1,\! \cdots\!, \phi_P\}$, with $\boldsymbol{\theta}\in \mathcal{T}^P$, where $\mathcal{T}^P$ is the domain of search.

\subsection{State-of-the-Art \ac{ML} Search}
\vspace{-1ex}

The \ac{SotA} parameter estimation method for \ac{MIMO}-\ac{OTFS} radar \cite{Gaudio_RadarConf20_2020} is based on a \ac{ML}-based search over the set of $4P$ parameters $\boldsymbol{\theta}$ described by
\vspace{-0.7ex}
\begin{equation}
\setlength{\abovedisplayskip}{2pt}
\setlength{\belowdisplayskip}{2pt}
\hat{{\boldsymbol{\theta}}} = \underset{\boldsymbol{\theta} \in \mathcal{T}^P}{\mathrm{argmin}} \; \Big| \mathbf{y} - \sum_{p=1}^{P} (h'_p \mathbf{G}_p) \mathbf{x} \Big|^2
\label{eq:sota_ML_sol},
\end{equation}
where $\mathbf{G}_p$ denotes $\mathbf{G}_p(\tau_p , \nu_p , \phi_p)$ for simplicity.

For a fixed set of $\phi_p,\tau_p,\nu_p$, the \ac{ML} estimate of $h'_p$ can be readily obtained as the solution of
\vspace{-0.5ex}
\begin{equation}
    \setlength{\abovedisplayskip}{2pt}
    \setlength{\belowdisplayskip}{2pt}
    \mathbf{x}\herm \mathbf{G}\herm_p \bigg(\sum_{q=0}^{P-1} h'_q \mathbf{G}_q \bigg)\mathbf{x} = \mathbf{x}\herm \mathbf{G}\herm_p \mathbf{y}, \;\; p = 0,\dots,P-1,
\label{eq:channel_coeff_update_eqn}    
\vspace{-0.5ex}
\end{equation}
which can be incorporated into eq. \eqref{eq:sota_ML_sol} to yield a reformulated maximization problem given by
\vspace{-0.5ex}
\begin{eqnarray}
\hat{{\boldsymbol{\theta}}} = \underset{\boldsymbol{\theta} \in \mathcal{T}^P}{\mathrm{argmax}} \; \Bigg( \sum_{p=1}^{P} \frac{||\mathbf{y}\herm \mathbf{G}_p \mathbf{x}||_2^2}{ ||\mathbf{G}_p \mathbf{x}||_2^2} - && \label{eq:sota_ML_sol_reformulated} \\[-0.5ex]
&&\hspace{-21ex} \sum_{p=1}^{P}\frac{(\mathbf{y}\herm \mathbf{G}_p \mathbf{x}) \cdot \mathbf{x}\herm (\mathbf{G}_p\herm \sum_{q\neq p} h'_q \mathbf{G}_q)\mathbf{x}}{||\mathbf{G}_p \mathbf{x}||_2^2} \Bigg) \!\nonumber,
\end{eqnarray}
where the two terms in the objective function respectively relate to the useful and interference signals for each target.

\ac{SotA} methods address the \ac{ML} optimization problem of equation \eqref{eq:sota_ML_sol_reformulated} utilizing a grid search-based method on the set of radar parameters $\{\tau_1,\!\cdots\!, \tau_P, \nu_1,\!\cdots\!,\nu_P, \phi_1,\!\cdots\!, \phi_P\} \!\in\! \{\mathbb{R}\! \times\! \mathbb{R}\! \times \![-\frac{\pi}{2},\frac{\pi}{2}]\}^P$.
However, although this method can accurately estimate the radar parameters at a desired precision\footnote{The precision of the continuous domain search can be determined by the spacing of the grid search, which can be iteratively reduced \cite{Gaudio_RadarConf20_2020}.}, it ultimately requires a search over the $3P$-dimensional space which is inefficient and impractical for large system sizes.

We therefore propose in the sequel a lower-complexity solution whereby \ac{AoA} estimation is performed first, independent of the other parameters, followed by the estimation of delay and Doppler via an \ac{LMMSE}-based \ac{ML} grid search.

\vspace{-3ex}
\subsection{Proposed Two-Step Radar Parameter Estimation}
\vspace{-1ex}

Let us start by recognizing that the channel model expressed by equation \eqref{eq:otfs_G} implies a separation between the \acp{AoA}, contained in the term $\mathbf{b}(\phi_p)\mathbf{a}\herm (\phi_p) \mathbf{f}_{\mathrm{BF}}$, and the delay and Doppler parameters, contained in the term $\mathbf{\Psi}^p \in \mathbb{C}^{N_\mathrm{a} NM\times NM}$, such that $\phi_p$ can be estimated independently of $\tau_p$ and $\nu_p$, as described below.

\vspace{-3ex}
\subsubsection{Root MUSIC-based \ac{AoA} Estimation}
\label{sec:AoAEstimation}
\vspace{-1ex}

While many frequency estimation methods exist, $e.g.$ \cite{Chen_Book_2010, Solak_TSP_2022}, for the sake of simplicity we consider the classic and efficient Root \ac{MUSIC}\cite{Ko_WSCE_2018} approach.

Let $\mathbf{U} \in \mathbb{C}^{N_a \times MN}$ be an unstacked and transposed reshaping of the received signal $\mathbf{y} \in \mathbb{C}^{N_a MN \times 1}$, such that the $q$-th set of $NM$ elements of $\mathbf{y}$ corresponds to the $q$-th row of $\mathbf{U}$ and consider the covariance matrix $\mathbf{R}_\text{UU}$
\begin{equation}
\mathbf{R}_\text{UU} \triangleq \mathbf{U} * \mathbf{U}\herm \;\;\;\; \in \mathbb{C}^{N_\mathrm{a} \times N_\mathrm{a}},
\label{eq:covariance_matrix}
\end{equation}

Since $\mathbf{R}_\text{UU}$ is Hermitian, its eigenvalues are all real, such that we may order its eigenvectors decreasingly, which shall be denoted $\{\mathbf{v_1 , v_2 , \dots , v_{\text{$N_a$}}}\}$.
Then, the eigenvectors $\{\mathbf{v_1 , \dots , v_{\text{p}}}\}$ span the signal subspace of $\mathbf{R}_\text{UU}$, while the remaining $N_a - p$ eigenvectors correspond to its noise subspace, which is orthogonal to the latter.
Denoting the noise subspace of $\mathbf{R}_\text{UU}$ by $\mathbf{V}_N$, and defining the matrix
\vspace{-0.5ex}
\begin{equation}
\mathbf{C} \triangleq \mathbf{V}_N * \mathbf{V}_N\herm \;\;\;\; \in \mathbb{C}^{N_\mathrm{a} \times N_\mathrm{a}},
\label{eq:c_hermitian}
\end{equation}
the \ac{AoA} of the $p$ targets can be obtained from the classical \ac{MUSIC} spectrum given by
\vspace{-0.5ex}
\begin{equation}
\textbf{P}(\phi) = \big(\big| \mathbf{a}(\phi)\herm \cdot \mathbf{C} \cdot \mathbf{a}(\phi) \big|\big)^{-1},
\label{eq:MUSIC_pseudospectrum}    
\end{equation}
or, alternatively, from the roots of the polynomial
\vspace{-0.5ex}
\begin{equation*}
\mathbf{a}(\phi)\herm \!\cdot\! \mathbf{C} \!\cdot\! \mathbf{a}(\phi) \!
= \!\!\!\sum_{m=1}^M \!\sum_{n=1}^N\! e^{-j(m-1)\pi \sin(\phi)}\!\! \cdot\! C_{mn} \!\cdot\! e^{j(n-1)\pi \sin(\phi)}
\end{equation*}
\begin{equation}
\hspace{6ex}= \hspace{-3ex}\sum_{i = -N_a + 1}^{N_a - 1} \!\!\!\!\!\! c_i \; e^{j \pi l \sin(\phi)}\!= \hspace{-3ex}\sum_{i = -N_a + 1}^{N_a - 1} \!\!\!\!\!\! c_i \; z^i\triangleq\textbf{D}(z),
\label{eq:polynomial_expression_final}
\vspace{-1ex}    
\end{equation}
where $c_i$ represents the sum of the diagonal elements of $\mathbf{C}$ and we implicitly defined $z \triangleq e^{-j \pi \sin(\phi)}$ and the \ac{MUSIC} polynomial $\textbf{D}(z)$.

Denoting the $p$-th root of $\textbf{D}(z)$ by $z_p$, the corresponding \ac{AoA} (in radians) of the $p$-th target is given by
\vspace{-0.5ex}
\begin{equation}
\hat{\phi}_p = -\sin^{-1} \Big( \frac{1}{\pi} \cdot \text{arg}(z_p) \Big).
\label{eq:final_AoA_estimate}    
\end{equation} 

Although the procedure concisely described above is very well known, it has not been proposed before (to the best of our knowledge) for the estimation of \ac{AoA} separately from delay and Doppler parameters, which is our actual contribution on the matter.
In possession of the estimates $\hat{\phi}_p$, we introduce another contribution consisting of a novel optimization problem for the estimation of the $\tau_p$ and $\nu_p$, which in addition to enjoying a search over a lower dimension also includes an improvement in combatting the effect of noise by means of the incorporating an \ac{LMMSE} filter.

\vspace{-2ex}
\subsubsection{\ac{LMMSE}-based \ac{ML} Estimation}
\label{sec:DDEstimation}
\vspace{-1ex}

\begin{subequations}
\label{eq:Theta_LMMSE_ML}
Straightforwardly, consider the reformulation of the minimization problem in eq. \eqref{eq:sota_ML_sol} incorporating robustness against noise by means of an \ac{LMMSE} procedure, namely
\vspace{-0.5ex}
\begin{equation}
\hat{{\boldsymbol{\theta}}}_\text{W} = \underset{\boldsymbol{\theta} \in \mathcal{T}^P}{\mathrm{argmin}} \; \big| \mathbf{x} - \mathbf{W}_\text{\ac{LMMSE}} \cdot \mathbf{y} \big|^2,
\label{eq:rreformulated_LMMSE_ML}    
\end{equation}
where
\vspace{-0.5ex}
\begin{equation}
\mathbf{W}_\text{\ac{LMMSE}} = \big( \mathbf{\Delta}\herm \mathbf{\Delta} + \sigma_w^2 \mathbf{I} \big)^{-1} \mathbf{\Delta}\herm,
\label{eq:W_LMMSE_definition}    
\end{equation}
with $\mathbf{\Delta} \triangleq \sum_{p=1}^{P} \big(h'_p \!\cdot \mathbf{G}_p(\tau_p , \nu_p , \phi_p)\big)$.
\end{subequations}

Notice, however, that the above \ac{LMMSE} cannot be applied directly for the estimation of all three radar parameters jointly, since in such a case the matrix $\mathbf{\Delta}$ would depend on the unknown \ac{AoA}, delay and Doppler parameters.
In contrast, thanks to the contribution of the step described earlier, the decoupled version of $\mathbf{\Delta}$ reduces to $\mathbf{\Psi}$, yielding the following new \ac{LMMSE}-based minimization problem over the reduced delay-Doppler  space
\begin{equation}
\hat{{\boldsymbol{\theta}}}_\text{2D} = \underset{(\tau, \nu)}{\mathrm{argmin}} \; \big| \mathbf{x} - \big( \mathbf{\Psi}\herm \mathbf{\Psi} + \sigma_w^2 \mathbf{I} \big)^{-1} \mathbf{\Psi}\herm \cdot \mathbf{y} \big|^2 .
\label{eq:reformulated_2D_LMMSE_ML}    
\end{equation}

Since we have information on $\mathbf{x}$, the structure of $\mathbf{\Psi}$ and y, we use eq. \eqref{eq:reformulated_2D_LMMSE_ML} to find the $P$ pairs $(\hat{\tau}_p, \hat{\nu}_p)$ which correspond to the $p$-th target parameters of each target. Note that some preprocessing is required for $\mathbf{y}$ since we have a MIMO system and therefore, the dimensionalities of $\mathbf{y}$ and $\mathbf{x}$ do not agree.

\vspace{-2ex}
\section{Performance Analysis}
\label{sec:perf_analysis}
\vspace{-2ex}

\subsection{Fundamental Limits}
\vspace{-1ex}

Before empirically assessing the efficacy of the proposed method, let us discuss a couple of fundamental limits on the \acp{RMSE} of the estimated parameters.
To that end, consider first the channel model in \eqref{eq:otfs_channel_eq} and define 
\vspace{-0.5ex}
\begin{equation}
s_p^{[n,m,t]} \!\triangleq\! A_p  e^{j\psi_p} b_t(\phi_p)a_t^*(\phi_p) f_t\! \sum_{k=0}^{L-1} \sum_{\ell=0}^{M\!-\!1}\! \mathbf{\Psi}_{n,k}^p[m,\ell] x_{k,\ell}, \!\!
\label{eq:CRLB_1}    
\end{equation}
\newpage

\noindent where $A_p = |h'_p|$ and $\psi_p = \angle (h'_p)$ denote the amplitude and phase of $h'_p$, respectively and $(n,m,t)$ denote time, subcarrier and antenna respectively. 

Then, the $(i,j)$-th element of the $5P \times 5P$ Fisher information matrix \cite{Gaudio_TWC_2020, Gaudio_RadarConf20_2020} corresponding to the estimate parameter vector $\hat{\boldsymbol{\theta}}$ is given by
\vspace{-0.5ex}
\begin{equation}
[\mathbf{I}(\boldsymbol{\theta})]_{i,j} = \frac{2}{N_0} \text{Re} \left\{ \sum_{n,m,t}  \left[  \frac{ \partial s_p^{[n,m,t]} } { \partial {\boldsymbol{\theta}}_i }  \right]^* \left[  \frac{ \partial s_q^{[n,m,t]} } { \partial {\boldsymbol{\theta}}_j }  \right] \right\},
\label{eq:CRLB_2}
\vspace{-0.25ex}
\end{equation}
from which the desired \acp{CRLB} can be computed as the diagonal elements of  $\mathbf{I}(\boldsymbol{\theta})^{-1}$. 

In the case of the delay and Doppler parameter, in addition to the \ac{CRLB}, we shall compare our simulation results also to the resolution limit of the corresponding search, taking into account also the number of refinements of the search grid \cite{Gaudio_TWC_2020, Gaudio_RadarConf20_2020}.

\vspace{-3ex}
\subsection{Simulation Setup}
\vspace{-1ex}

For simplicity, we consider in our simulations a mono-static \ac{MIMO} base station and a single target, with $N_a = 16$, $N = 16$, $M = 16$, $f_c = 60 \; \text{GHz}$, $B = 150 \; \text{MHz}$ and \ac{OTFS} signals build over QAM symbols.
It is assumed that the target is located at a distance $r = 14 \;\text{m}$ and traveling with a velocity of $v = 60 \;\text{km/h}$ directly towards the base station. 

The results are shown as a function of the radar \ac{SNR} of the reflected signal, defined as \cite{Gaudio_RadarConf20_2020} 
\begin{equation}
\text{SNR}_{\text{rad}} \triangleq \frac{\lambda^2 \sigma_{\text{rcs}} G^2}{(4\pi)^3 r^4} \frac{P_{\text{avg}}}{\sigma_w^2}, 
\label{eq:radar_SNR}
\end{equation}
where $\lambda = c/f_c$ is the wavelength , $c$ is the speed of light, $\sigma_{\text{rcs}}$ is the radar cross-section of the target in $\text{m}^2$ (we set $\sigma_{\text{rcs}} = 1$), $r$ is the distance between the transmitter and receiver, $P_{\text{avg}} = 1$, and $\sigma_w^2$ is the variance of the AWGN noise.

\vspace{-3ex}
\subsection{Simulation Results}
\vspace{-1ex}

\begin{figure}[H]
\centering
\includegraphics[width=\columnwidth]{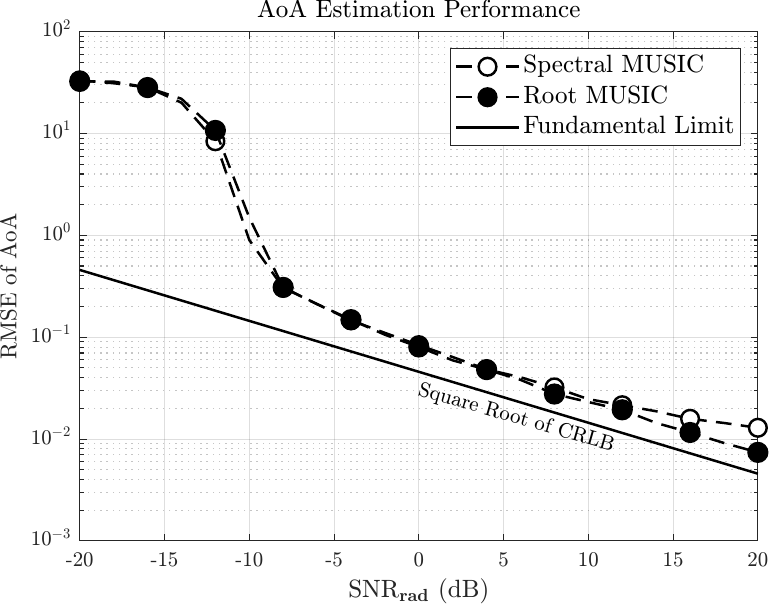}
\vspace{-4ex}
\caption{Performance of decoupled \ac{AoA} estimation via \ac{MUSIC}.}
\label{fig:AoAEstimation}
\end{figure}

First, we show in Fig. \ref{fig:AoAEstimation} the performance of the first step described in Subsection \ref{sec:AoAEstimation}, whereby \ac{AoA} is estimated via Spectral and Root \ac{MUSIC}.
The fact that the results show the usual good performance associated with subspace-based methods \cite{Chen_Book_2010} indicates that the decoupling of \ac{AoA} and \ac{DD} estimation does not result in any penalty in terms of accuracy, in spite of the complexity reduction, from $\mathcal{O}\big((N_a \cdot N \cdot M)^3\big)$ for \ac{SotA} schemes to $\mathcal{O}\big(N_a^3+ (NM)^3+(N_aNM)^2\big)$ of the proposed method\footnote{Derivations are omitted due to page limitations.}.
\vspace{-2ex}

\begin{figure}[H]
\begin{subfigure}[b]{\columnwidth}
\centering
\includegraphics[width=\columnwidth]{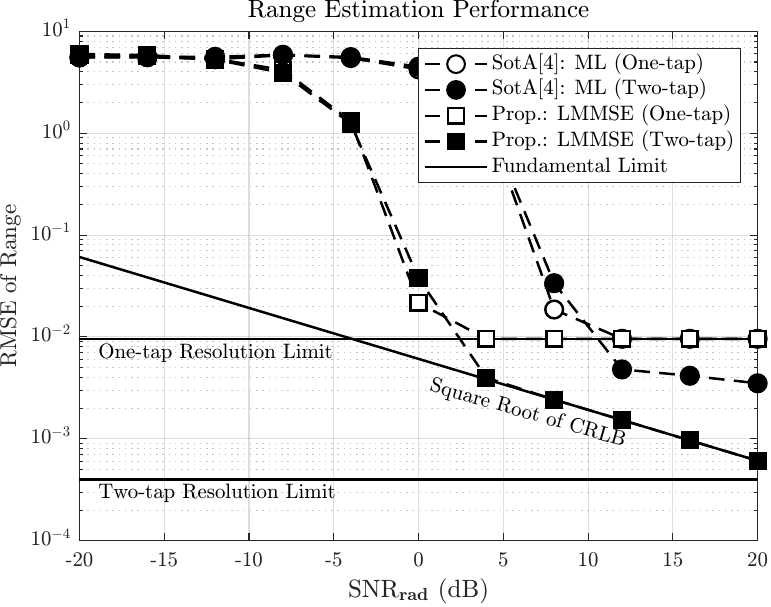}
\label{fig:labelx}
\vspace{-2ex}
\end{subfigure}
\begin{subfigure}[b]{\columnwidth}
\centering
\includegraphics[width=\columnwidth]{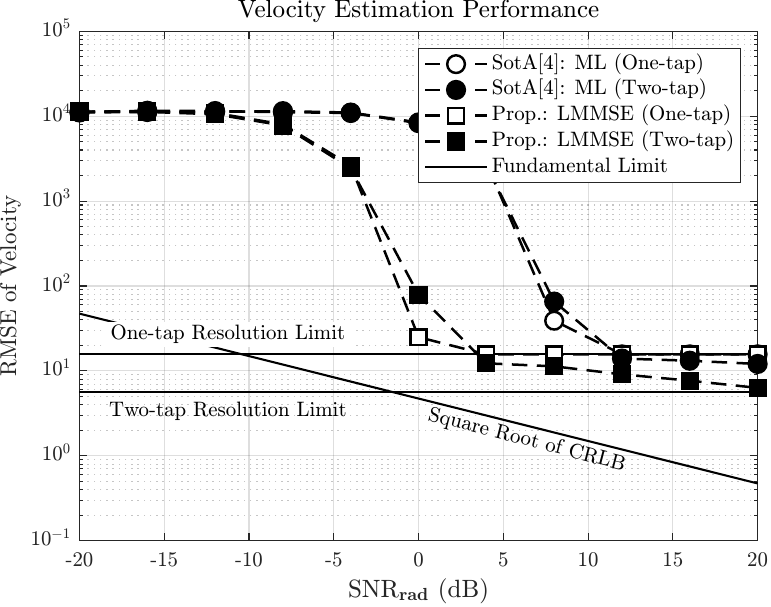}
\label{fig:labelxx}
\end{subfigure}
\vspace{-6ex}
\caption{Performance of delay and Doppler estimation via \ac{LMMSE}-improved \ac{ML} search.}
\label{fig:main_fig}
\vspace{-2ex}
\end{figure}

Next, we compare in Fig. \ref{fig:main_fig} the performances of the proposed and \ac{SotA} \cite{Gaudio_RadarConf20_2020} methods.
It is found that the proposed method significantly improves estimation performance, especially at lower radar \acp{SNR}, to the extent that with a one-tap \ac{ML} search, the \ac{RMSE} of the proposed method in the range from -5 dB to 0 dB is similar to that of the \ac{SotA} alternative in the range from 0 dB to 10 dB, indicating a gain of 10 dB in that region.
In addition, it is also found that for higher \acp{SNR}, the two-tap \ac{ML} search with the proposed method approaches the \ac{CRLB} faster than the \ac{SotA} scheme.

\vspace{-2ex}
\section{Conclusion}
\label{sec:conclusion}
\vspace{-2ex}
We proposed a novel two-step algorithm to estimate the radar parameters from \ac{OTFS} signals sequentially, with the \ac{AoA} estimates obtained decoupled from range/velocity.
Thanks to the approach, in addition to the resulting lower complexity due to the reduced \ac{2D} grid search in the \ac{DD} domain, the decoupling of \ac{AoA} and \ac{DD} estimation enables the incorporation of an \ac{LMMSE} procedure into the \ac{ML} estimation of range and velocity, which were shown via simulations to significantly outperform the \ac{SotA} and approaches the fundamental limits.

\vfill\pagebreak

\bibliographystyle{IEEEbib}
\bibliography{strings,refs}

\end{document}